\documentclass[twocolumn]{aastex61}

\def\gsim{ \lower .75ex \hbox{$\sim$} \llap{\raise .27ex \hbox{$>$}} }
\def\lsim{ \lower .75ex\hbox{$\sim$} \llap{\raise .27ex \hbox{$<$}} }

\def\sc{Schwarzschild}
\def\beq{\begin{equation}}
\def\eeq{\end{equation}}

\def\sc{Schwarzschild}

\usepackage[utf8]{inputenc}
\usepackage{amssymb,amsmath,psfig,times}
\usepackage{natbib}

\received{\today}
\revised{xxx}
\accepted{xxx}
\submitjournal{ApJL}

\shorttitle{\emph{Fermi} transient J1544-0649: a flaring radio-weak BL Lac}
\shortauthors{Bruni et al.}
\begin{document}

\title{\emph{Fermi} transient J1544-0649: a flaring radio-weak BL Lac}

\correspondingauthor{Gabriele Bruni}
\email{gabriele.bruni@iaps.inaf.it}

\author[0000-0002-5182-6289]{Gabriele Bruni}
\affil{INAF - Istituto di Astrofisica e Planetologia Spaziali, via Fosso del Cavaliere 100, I-00133 Roma, Italy}

\author{Francesca Panessa}
\affiliation{INAF - Istituto di Astrofisica e Planetologia Spaziali, via Fosso del Cavaliere 100, I-00133 Roma, Italy}

\author{Gabriele Ghisellini}
\affiliation{INAF - Osservatorio Astronomico di Brera, via E. Bianchi 46, I-23807 Merate, Italy}

\author{Vahram Chavushyan}
\affiliation{INAOE - Instituto Nacional de Astrof\'isica \'Optica y Electr\'onica, Apartado Postal 51-216, 72000 Puebla, Mexico}

\author{Harold A. Pe\~{n}a-Herazo}
\affiliation{INAOE - Instituto Nacional de Astrof\'isica \'Optica y Electr\'onica, Apartado Postal 51-216, 72000 Puebla, Mexico}
\affiliation{Università degli Studi di Torino - Dipartimento di Fisica, via Pietro Giuria 1, 10125 Torino, Italy}
\affiliation{INFN - Istituto Nazionale di Fisica Nucleare, Sezione di Torino, 10125 Torino, Italy}

\author{Lorena Hern\'andez-Garc\'ia}
\affiliation{IFA - Instituto de F\'isica y Astronom\'ia, Facultad de Ciencias, Universidad de Valpara\'iso, Gran Breta\~{n}a 1111, Playa Ancha, Valpara\'iso, Chile }

\author{Angela Bazzano}
\affiliation{INAF - Istituto di Astrofisica e Planetologia Spaziali, via Fosso del Cavaliere 100, I-00133 Roma, Italy}

\author{Pietro Ubertini}
\affiliation{INAF - Istituto di Astrofisica e Planetologia Spaziali, via Fosso del Cavaliere 100, I-00133 Roma, Italy}

\author{Alex Kraus}
\affiliation{MPIfR - Max Planck Institute for Radio Astronomy, auf dem H\"ugel 69, D-53121 Bonn, Germany}



\begin{abstract}
On May 15th, 2017, the \emph{FERMI}/LAT gamma-ray telescope observed a transient source not present in any previous high-energy catalogue: J1544-0649. It was visible for two consecutive weeks, with a flux peak on May 21st. Subsequently observed by a \emph{Swift}/XRT follow-up starting on May 26, the X-ray counterpart position was coincident with the optical transient ASASSN-17gs = AT2017egv, detected on May 25, with a potential host galaxy at $z$=0.171. %In radio band, the transient corresponds to NVSS J154419-064913, with a flux density at 1.4 GHz of 46.6 mJy. The flat radio SED from literature data suggested a radio-weak BL Lac source, since the radio power at 1.4 GHz is lower than the radio-loud threshold.\\
We conducted a 4-months follow-up in radio (Effelsberg-100m) and optical (San Pedro M\'artir, 2.1m) bands, in order to build the overall Spectral Energy Distribution (SED) of this object. The radio data from 5 to 15 GHz confirmed the flat spectrum of the source, favoring a line of sight close to jet axis, not showing significant variability in the explored post-burst time-window. The Rx ratio, common indicator of radio loudness, gives a value at the border between the radio-loud and radio-quiet AGN populations. The Ca$_{\rm{II}}$ H\&K break value (0.29$\pm$0.05) is compatible with the range expected for the long-sought intermediate population between BL Lacs and FRI radio galaxies. An overall SED fitting from Radio to $\gamma$-ray band shows properties typical of a low-power BL Lac. As a whole, these results suggest that this transient could well be a new example of the recently discovered class of radio-weak BL Lac, showing for the first time a flare in the gamma/X-ray bands.
\end{abstract}




\section{Introduction} \label{sec:intro}

The release of the \emph{Fermi} mission catalogues has revealed several unidentified gamma-ray sources, whose association with lower frequencies emission is still under study. Among these, radio-weak BL Lac objects (\citealt{Massaro17} and references therein) are challenging our comprehension of the radio-loud phase in active galactic nuclei (AGN). These extragalactic objects present the same optical properties of BL Lacs - that together with flat spectrum radio quasars define the Blazar class - but radio weak (mostly undetected in major radio surveys). They behave differently from what expected for this kind of objects, normally used as calibrators for radio observatories (e.g. 3C\,286, 3C\,273, BL Lac, OJ\,287). Until now, only a handful of radio-weak BL Lacs have been presented in the literature, and their collocation in the context of the more general unified model of AGN (Urry \& Padovani 1995) is still under debate. Indeed, understanding the radio-weak (radio-quiet) AGN population is of crucial importance in order to unveil the conditions that trigger the launching and collimation of relativistic jets, and their relation with other constituents of AGN such as relativistic outflows. Most recent observational results (\citealt{Boccardi16}) suggest that the jet base in some sources is wide enough to favour the \cite{Blandford} model, in which the jet can be originated from the accretion disk: BL Lacs are believed to be the low-accreting fraction of the Blazar class, at the end of an evolutionary track starting from high-accreting flat spectrum radio quasar (\citealt{Cavaliere}).  Thus, radio-weak BL Lacs are the ideal objects to probe the minimum conditions needed to form a jet. Their existence and fraction would also have an impact on the gamma-ray association with radio counterparts, currently based on major radio surveys like FIRST (\citealt{Becker}) and NVSS (\citealt{Condon}), that basically do not include the radio-quiet population  lying below the $\sim$1 mJy flux density threshold.

\begin{figure}
\centering
\includegraphics[width=1.0\linewidth]{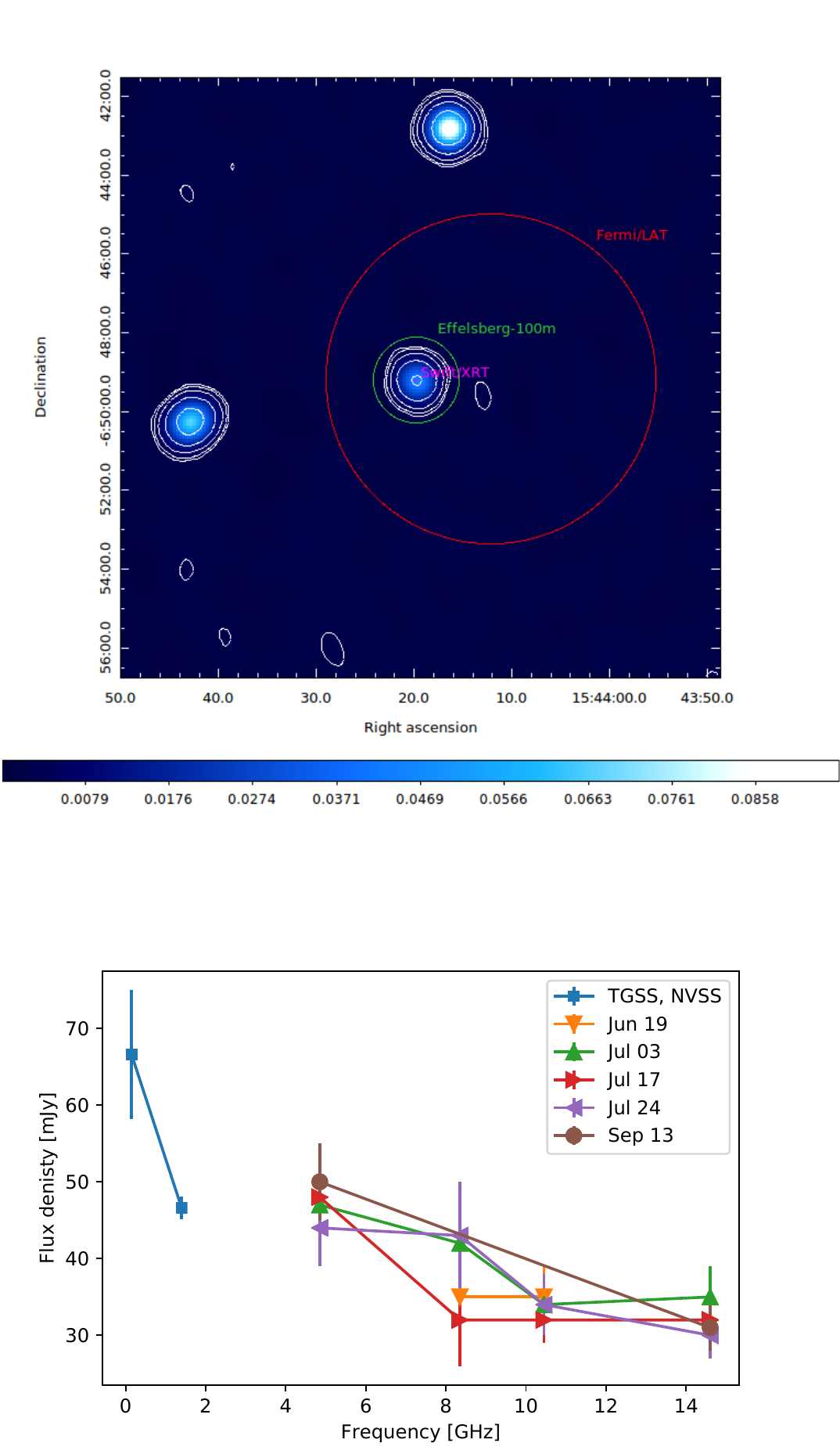}
%\caption{A mouse}
\caption{Top panel: NVSS map centered on source J154419-064913. Angular resolution and target positions for Effelsberg-100m ($\sim$2.5 arcmin at 4.85 GHz), \emph{Swift}/XRT (90\% containment radius of 1.4 arcsec) and \emph{Fermi}/LAT (95\% containment radius of 0.07 deg) are shown. Flux density scale from NVSS on bottom is in Jy/beam, contours are in logarithmic scale from 3$\sigma$ to 1000$\sigma$. Bottom panel: Effelsberg-100m radio SED at different epochs (post-burst), together with flux densities from the literature (TGSS and NVSS, pre-burst).  \label{fig1}}
\end{figure}

On May 15th (2017) \emph{Fermi}/LAT detected a transient, not associated with any known gamma-ray source. Named as \emph{Fermi} J1544-0649, it showed clear detection for two consecutive weeks, with a peak on May 21. It was reported by \cite{Ciprini17} on June 11, together with a follow-up observation by \emph{Swift}/XRT, measuring an enhancement in X-ray emission at a position corresponding to the optical transient detected on May 25 (ASASSN-17gs = \href{https://wis-tns.weizmann.ac.il/object/2017egv}{AT2017egv}). The host galaxy has been suggested to be 2MASX J15441967-0649156, observed in optical band as follow-up of the \emph{Fermi} transient, and for which a redshift of $z=0.171$ has been estimated (\citealt{Chornock}). The source is also present in the NVSS catalogue (\citealt{Condon}) with a flux density at 1.4 GHz of 46.6 mJy, and a marginally resolved morphology of about 2' (see Fig. \ref{fig1}). More recently, it has been detected by the TGSS survey at 150 MHz (\citealt{Intema17}), performed with the GMRT, with a flux density of 67 mJy. The flat radio spectrum between these two measurements seems to suggest a Blazar-like emission, with the jet oriented towards the observer, but with a power ascribable to radio-quiet AGN (see Sec. 2 for a detailed discussion). Indeed, it has a $L_{1.4\,GHz}=3.9\times10^{31} {\rm{erg/s/Hz}}$, lower than the threshold limit definition ($L_{1.4\,GHz}=10^{32.5} {\rm{erg/s/Hz}}$, \citealt{Gregg96}). This makes this transient a possible candidate for an AGN class still under debate: radio-weak BL Lacs.

\section{Observations and archive data} \label{sec:style}
%In the following we describe the observations, in radio and optical bands, collected after the transient detection by \emph{Fermi}, and data gathered from archives.

\subsection{Effelsberg-100m single dish data}

\begin{figure*}
\centering
\includegraphics[width=1.0\linewidth]{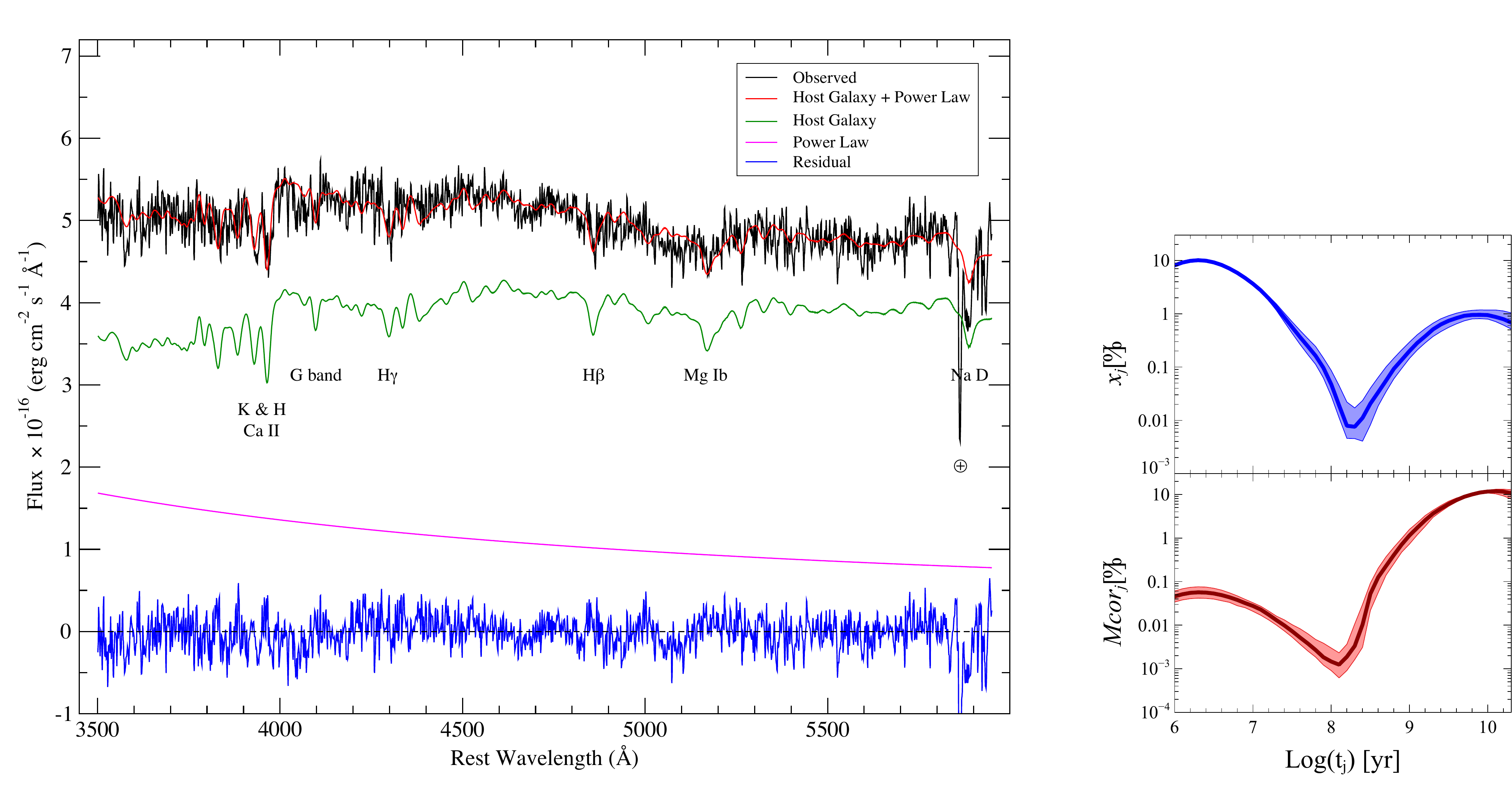}
%\caption{A mouse}
\caption{Left panel: spectral synthesis for the object J154419-064913, performed by {\tt STARLIGHT} using 150 SSP and six power laws to simulate the AGN continuum emission. The observed spectra, host galaxy model, AGN component and residuals are shown. Right panel: population mixture results obtained for 100 realizations with STARLIGHT. (Top-panel) The median curve of the contribution of the light-weighted population vector $x_{j}$ (blue) and (Bottom-panel) median curve of the contribution of the mass-weighted population vector $Mcor_{j}$. Both population vectors are in percentage and logarithmic scale and plotted against each SSP with age $t_{j}$. In both panels, the shadow regions represent first and third quartiles. \label{fig2}}
\end{figure*}

We obtained a ToO observations with the Effelsberg-100m single dish radio telescope to monitor the radio SED of our target in a $\sim$4-months time window after the \cite{Ciprini17} discovery. Observations were carried out in cross-scan mode at 4.85, 8.35, 10.45, and 15 GHz, in order to cover the entire cm-band. The number of subscans was 16 at 4.85 GHz and 8.35 GHz, and 32 at 10.45 GHz and 15 GHz, reaching a typical RMS of $\sim$2 mJy and $\sim$1 mJy, respectively. Pointing/focussing on convenient nearby calibrators were performed before each observing slot. The flux density scale was calibrated using multiple scans on 3C\,286 for each epoch, taking as reference the \cite{Baars} scale: given the variable weather conditions, the flux scale uncertainty can be considered $\sim$10\% of the total flux density measurement. Data were reduced in {\tt{TOOLBOX}}\footnote{\href{https://eff100mwiki.mpifr-bonn.mpg.de/doku.php}{https://eff100mwiki.mpifr-bonn.mpg.de/doku.php}}, extracting the flux densities via Gaussian fit of the cross-scans. The final error has been calculated via quadratic sum of the cross-scan RMS and flux scale error.

In total, we collected 5 epochs at $t$=+35, +49, +63, +70, +121 days after the initial burst detected by \emph{Fermi}. Measurements are reported in table \ref{Eff_fluxes}, and plotted in figure \ref{fig1}: values at the same frequency from different epochs are compatible within errors, not indicating radio variability up to four months after the burst in the gamma-ray band. A correlation between gamma-ray and radio emission in blazars has been found by different authors (\citealt{Schinzel, CasadioCTA, Chidiac, Karamanavis, Lisakov}) with time-lags spanning between days to months for the radio associated emission, and flux density enhancements of a factor of 2 or more. In this case, the low activity of the source could further expand the delay between detection in the different bands, being more than the 4 months explored here. Another possibility is that the lower energy involved with respect to classic blazars produces a variability magnitude smaller than our error bars from Effelsberg-100m measurements ($<$5 mJy, i.e. $<$10\% of flux density).

The spectral index ($\alpha$, adopting the convention $S=\nu^\alpha$) calculated between 4.85 GHz and 14.60 GHz shows values $>-0.5$ for all epochs, suggesting an orientation of the jet near the line of sight, typical of blazars. This confirms the estimate at lower frequencies using pre-burst flux densities from the literature, giving an $\alpha=-0.16\pm0.21$ between 150 MHz (TGSS, 66.6$\pm$8.4 mJy) and 1.4 GHz (NVSS, 46.6$\pm$1.5 mJy).

In figure \ref{fig1} we show the NVSS map at 1.4 GHz of our target, with an angular resolution of 45$\times$45 arcsec. Other two intervening sources are visible, but not related with the emission of the transient. The Effelsberg-100m angular resolution and pointing position at the lower frequency observed (4.85 GHz) is reported for comparison, together with the \emph{Fermi}/LAT and \emph{Swift}/XRT 95\% and 90\% containment radius, respectively, from the corresponding ATels. The identification of the \emph{Swift}/XRT transient with the NVSS source is evident, and the lack of other radio counterparts in the {\emph{Fermi}} and Effelsberg-100m beams results in a bona-fide association as well. The NVSS map itself shows a marginally resolved structure, with deconvolved linear size of 15$\times$6 arcsec, and a position angle of 24 degrees. At the spatial scale of the source (2.912 kpc/arcsec, for a flat Universe with $\Omega_{\rm{m}}=0.3$, $\Omega_{\Lambda}=0.7$, and $H_0=70$ km/s/Mpc - Planck Collaboration XVI 2014), this translates into a projected size of 44$\times$17 kpc, comparable with the dimension of extended jets from common blazars (e.g. 3C273, BL Lac).

In order to estimate the radio-loudness of the source with a further indicator (in addition to the \citealt{Gregg96} mentioned above), we adopted $R_{X}$ (\citealt{Terashima}). For this calculation we used the post-burst data at 5 GHz from our Effelsberg-100m campaign, and the flux from \emph{Swift}/XRT published in ATel \#10642. We obtained a value of $Log({\rm{Rx}})=-4.3$, near to the limit of $-4.5$ considered by \cite{Terashima} to divide the radio-loud and radio-quiet populations. This value is also in the range obtained for Seyfert galaxies by the same authors.

\begin{deluxetable}{ccccccc}
%\renewcommand\tabcolsep{4pt}
% \centering
% \begin{minipage}{152mm}
  \tablecaption{Radio flux densities collected with the Effelsberg-100m telescope for the 5 post-burst epochs in 2017. Days elapsed after burst are in column 2. Frequencies are in GHz, flux densities in mJy (columns 3-6). The last column reports spectral indices between 4.85 and 14.60 GHz.  \label{Eff_fluxes}}
%\scalebox{0.85}{
%\begin{tabular}{cccccc}
\tablehead{\colhead{Epoch} & Time & \colhead{4.85} & \colhead{8.35} & \colhead{10.45} & \colhead{14.60} & \colhead{$\alpha_{4.85}^{14.60}$}}
%\hline
%  \multicolumn{1}{c}{Epoch} &
%  \multicolumn{1}{c}{4.85}  &
%  \multicolumn{1}{c}{8.35}  &
%  \multicolumn{1}{c}{10.45} &
%  \multicolumn{1}{c}{14.60} &
%  \multicolumn{1}{c}{$\alpha_{4.85}^{14.60}$} \\  
%\hline
\startdata
  Jun 19	  &	+35		&	-    		  &  35$\pm$6  &  35$\pm$4  &  -	    &	-	\\
  Jul 3	  	  &	+49		&	47$\pm$5	  &  42$\pm$5  &  34$\pm$4  &  35$\pm$4 &	--0.27$\pm$0.25	\\
  Jul 17	  &	+63		&	48$\pm$5	  &  32$\pm$6  &  32$\pm$3  &  32$\pm$3 &	--0.37$\pm$0.23	\\
  Jul 24	  &	+70		&	44$\pm$5	  &  43$\pm$7  &  34$\pm$4  &  30$\pm$3 &	--0.35$\pm$0.25	\\
  Sep 13	  &	+121	&	50$\pm$5	  &  -	       &  -	        &  31$\pm$3 &	--0.43$\pm$0.23	\\
%\hline
%\end{tabular}
\enddata
\end{deluxetable}

\subsection{Optical observations}

%\begin{figure}
%\centering
%\includegraphics[width=1.0\linewidth]{./ASASSN-17gs_sp.pdf}
%\caption{A mouse}
%\caption{Spectral synthesis for the object J154419-064913, performed by {\tt STARLIGHT} using 150 SSP and six power laws to simulate the AGN continuum emission. The observed spectra, host galaxy model, AGN component and residuals are shown.\label{Spectrum}}
%\end{figure}

Optical spectra were obtained with the 2.1-m telescope of the Observatorio Astronómico Nacional at San Pedro Mártir (OAN-SMP), Baja California, México, on August 23rd 2017 (start at 03:30 UT). Clear sky conditions were present, with a seeing of 2.1 arc-seconds. The Boller \& Chivens spectrograph was tuned to the 3800  \AA~ to 8000 \AA~ range (grating 300 l/mm), with a spectral dispersion of 4.5 \AA/pix, corresponding to 10 \AA~ full width at half maximum (FWHM), derived from the FWHM of different emission lines of the arc-lamp spectrum. A 2.5 arcsecond slit was used. To calibrate the spectral measurements, the spectrophotometric standard star Feige 110 was observed during night.

The data reduction was carried out with the {\tt IRAF}\footnote{{\tt IRAF} is distributed by the National Optical Astronomy Observatories operated by the Association of Universities for Research in Astronomy, Inc. under cooperative agreement with the National Science Foundation.} software following standard procedures. The spectra were bias-subtracted and corrected with dome flat-field frames. Cosmic rays were removed interactively from all images. Arc-lamp (CuHeNeAr) exposures were used for the wavelength calibration. A spline function was fitted to determine the dispersion function (wavelength-to-pixel correspondence). Sky emission lines located at known wavelengths were removed during the calibration in wavelength. The obtained spectrum is shown in Fig. \ref{fig2}.

\subsubsection{Black hole mass estimate}

To evaluate the central black hole mass, we used the stellar population synthesis code {\tt STARLIGHT}. A detailed description of the {\tt STARLIGHT} code can be found in the publications of the Semi Empirical Analysis of Galaxies (SEAGal) collaboration (\citealt{Cid05, Cid07}). Before running {\tt STARLIGHT} we carried out the pre-processing steps required for the code implementation. First, the spectra were corrected for Galactic extinction assuming an E(B-V) value of 0.156, as computed by \cite{Schlegel}. Extinction in the galaxy is taken into account in the synthesis, assuming that it arises from a foreground screen with the extinction law of \cite{Cardelli}. 
We derived information about the stellar populations of the host galaxy in the STARLIGHT analysis. To simplify the analysis we decided to define the intervals of young ($t_{j}<10^{8}$ yr), intermediate ($10^{8}\le t_{j}\le 10^{9}$ yr) and old ($t_{j}>10^{9}$ yr) stellar populations as stated in Cid Fernandes et al. 2005.
The best fit is a combination of 150 single stellar populations (SSP) from the evolutionary synthesis models of \cite{Bruzual} and 6 power laws to represent the AGN continuum emission (see Fig. \ref{fig2} - e.g., \citealt{Leon}). Since BL Lacs are characterized by strong non-thermal emission at all frequencies from compact components with power-law continua, spectral indices of -0.5, -1.0, -1.5, -2.0, -2.5, and -3.0 were used to represent non-thermal jet emission typically observed in these AGN. As we can observe in Fig. \ref{fig2} , the light-weighted population vector is dominated  ($\sim$90\%) by a young stellar population although the contribution of this population to the mass-weighted population vector is less than 1\%, which can be explained by a recent star formation event. Moreover, we did not observe a significant contribution of the intermediate stellar population. In contrast, the old stellar population represents nearly 97\% of the mass-weighted population vector $Mcor_{j}$ and 9\%, of the light-weighted population vector, in agreement with the elliptical nature of the host galaxy of BL Lacertae objects.

We estimate the $M_{BH}$ from the velocity dispersion derived with {\tt STARLIGHT}. Assuming an average instrumental resolution, from the FWHM of the sky lines, of 18.5 \AA~ in the range between 4000-7000 \AA, this yields a corrected stellar velocity dispersion of 252$\pm$43 [km/s]. Using the $M-\sigma$ relation in \cite{Tremaine} we estimate a black hole mass of $M_{BH}$ = 3.4$\pm$1.4 $\times10^{8}$ [\(\textup{M}_\odot\)].

\subsubsection{Ca$_{\rm II}$ H\&K break}

To assess the spectral classification of our target, we measured the Ca$_{\rm II}$ H\&K break from optical spectrum (\citealt{Landt}). The value of the Ca$_{\rm II}$ break of C = 0.29$\pm$0.05 ensures that the galaxy is dominated by the thermal spectrum of the host rather than the non-thermal spectrum of an active nucleus or a relativistic jet. \cite{Landt} find that Ca$_{\rm II}$ H\&K break values $0.25<C<0.4$ could represent the long-sought population intermediate between the 'classical' BL Lacs and the FRI radio galaxies. Still, the same authors suggest that a value of 0.35 can be assumed to separate BL Lacs and radio galaxies, that makes our object a member of the first class.

%\subsection{INTEGRAL archive data}
%We used archive data from the INTEGRAL mission to verify the source emission in the soft gamma-ray band. [...]

\subsection{Multi-wavelength SED fitting}

%We made use of the ASI SSDC Sky Explorer\footnote{\href{https://tools.asdc.asi.it/SED/}{https://tools.asdc.asi.it/SED/}} to build the multi-wavelength SED, from radio to gamma-ray, from pre-burst archival data. Only data from radio to UV band were available (see Fig. \ref{SED}), namely from NVSS, WISE, and GALEX. 

%In addition to that, we added an upper limit from {\emph{INTEGRAL/ISGRI}} measurements of the field corresponding to our target position, resulting from the first 1000 orbits observations (Bird et al. 2016): the upper limit corresponds to a flux $<3.8\times10^{-12}$ [erg/cm$^2$/s] in the 20-40 KeV band, considering data collected from 2002 to 2012. 

%Sensitivity curves from {\emph{INTEGRAL/ISGRI}}, {\emph{INTEGRAL/PICsIT}}, {\emph{Swift/BAT}}, and {\emph{Fermi/LAT}} 4-years catalogue are also plotted: the overall faintness of the source is evident, making hard a detection in the soft gamma-ray band for {\emph{INTEGRAL}}.

%We then considered the after-burst detections from {\emph{Fermi/LAT}} (ATel \#10482), {\emph{Swift/XRT}} (ATel \#10642), plus Effelsberg-100m and San Pedro M\'artir photometry in radio and optical bands from our campaign, on order to compare with previous measurements [...]

%\begin{figure}
%\centering
%\includegraphics[width=1.0\linewidth]{./SED.png}
%\caption{A mouse}
%\caption{Multi-wavelength SED built from pre-burst archive data. Sensitivity curves for %{\emph{INTEGRAL/ISGRI}}, {\emph{INTEGRAL/PICsIT}}, {\emph{Swift/BAT}}, and {\emph{Fermi/LAT}} 4-years catalogue are shown. \label{SED}}
%\end{figure}

In order to compile a pre-burst multi-wavelength SED, we collected the available archive data with the ASI Science Data Center (ASDC) online tool\footnote{\href{http://tools.asdc.asi.it/}{http://tools.asdc.asi.it/}}. In addition to that, we considered the upper limit from {\emph{INTEGRAL/ISGRI}} measurements of the field corresponding to our target position, resulting from the first 1000 orbits observations (\citealt{Bird16}) in the 20-40 KeV band ($<3.8\times10^{-12}$ [erg/cm$^2$/s]). As post-burst data, we used the second Effelsberg-100m epoch (for which 4 frequencies were available), the \emph{Swift}/UVOT data corresponding to the first \emph{Swift}/XRT follow-up, and obviously the \emph{Swift}/XRT spectrum itself together with the \emph{Fermi} one.

\begin{figure} 
%\vskip -0.6 cm
%\hskip -0.4 cm
%\psfig{file=1544_mkn501l.pdf,width=10cm,height=9cm } 
\includegraphics[width=1.0\linewidth]{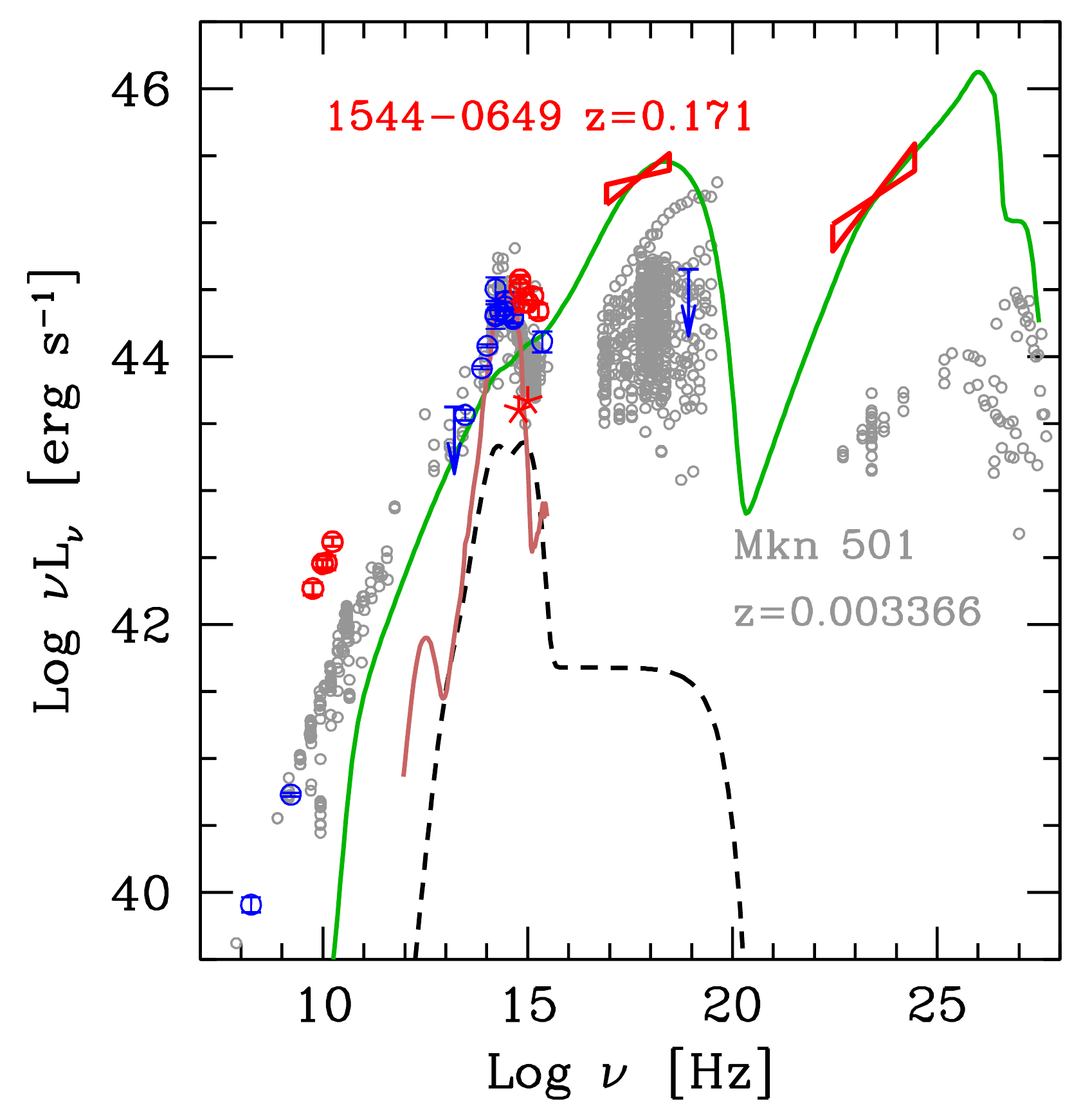}
%\vskip -0.5 cm
\caption{
The SED of 1544--0649 (blue and red symbols)
compared with archival of data of Mkn 501 (grey circles) taken from
ASDC. 
The red points correspond to the flare and post--flare emission,
blue points correspond the pre--flare epoch.
The solid green lines refer to the theoretical model while the
red solid line corresponds to a template for a typical host galaxy of BL Lac objects
(\citealt{Silva}).
The dashed black line is the spectrum of an accretion disk, together with
the IR torus radiation and the X--ray produced in the corona.
The two--humps in $\gamma$--rays correspond to the peak of the first 
and second order Compton scattering.
}
\label{sed}
\end{figure}

The overall spectral energy distribution of 1544--9649
shows the typical two--hump shape of blazars, commonly interpreted
by a synchrotron and inverse Compton emission from a relativistic jet.
Fig. \ref{sed} shows the overall SED, and compares it with the one
of Mkn 501, a typical low power, high energy peaked BL Lac object.
In both objects the host galaxy emission dominates the optical luminosity,
both have a rising (in $\nu F_\nu$) spectrum in X--rays and in $\gamma$--rays
and both show a large amplitude variability.
These similarities suggest that 1544--0649 is a ``blue" BL Lac object,
but more luminous (in its high state) than Mkn 501.

To the aim of reproducing the overall SED, we use the model
fully described in \cite{Ghisellini09}.
This model assumes that most radiation is produced by relativistic electrons
located at a distance $R_{\rm diss}$ from the black hole.
The emitting plasma is moving at a relativistic velocity $\beta c$
corresponding to the bulk Lorentz factor $\Gamma$.
The viewing angle $\theta_{\rm v}$ is small, so that 
the relativistic Doppler factor is 
$\delta\equiv 1/[\Gamma(1-\beta\cos\theta_{\rm v}]>1$.

% --------------------------------------
% --------------------------------------

The emitting particle energy distribution is derived 
solving a continuity equation accounting for  
radiative cooling and electron--positron pair production
and the contribution of these pairs to the observed flux.

The emission processes are synchrotron, synchrotron self--Compton (SSC)
and inverse Compton scattering off photons produced externally to the jet (EC),
by the disk, by the broad line region and by the molecular torus.

To account for the fast variability observed in BL Lacs, the source is generally assumed as compact, located at a $R_{\rm diss}=$900  \sc\ radii, and having a size ten times less (we are assuming that the jet is conical with a semi--aperture angle $\psi=5^\circ$).
The compactness of the source implies that the synchrotron spectrum is self--absorbed
below $\sim$100 GHz, and therefore cannot account for the radio flux at smaller frequencies.
This has to be produced by other, more extended, portions of the jet.

\subsubsection{SED properties}

Fig. \ref{sed} shows the overall SED of 1544--0649 together with 
our model, and Tab. \ref{para} lists the used parameters.
The SED of the host galaxy (red solid line) is obtained using the template 
of a giant elliptical galaxy from \cite{Silva}, 
while for the optical emission of the jet we have assumed
the decomposition shown in Fig. 3.

The model can capture some of the main features of the SED, but it
fails to well reproduce the shape of the $\gamma$--ray emission and 
the radio flux.
As mentioned, the latter can be produced by other, less compact, regions of the jet.
The disagreement with the $\gamma$--ray slope can instead be due to some
reprocessing of the emission.
At a $z=0.171$, in fact, the IR background can absorb part of the high energy radiation,
producing electron--positron pairs, that in turn interact with the cosmic microwave background
through the inverse Compton process, producing photons that are still $\gamma$--rays,
but of smaller energies. 
This process could then soften the observed $\gamma$--ray slope.

The parameters used for the model (see table \ref{para}) are rather typical for low power BL Lac objects,
whose SED peaks at X--ray energies (synchrotron) and in the TeV band (inverse Compton).

%------------------------------------------------
\begin{deluxetable*}{cccccccccccc}
\tablecaption{
Adopted parameters for the jet models.
Col. [1]: redshift;
Col. [2]: black hole mass in solar masses;
Col. [3]: disk luminosity in units of $10^{45}$ erg s$^{-1}$;
Col. [4]: disk luminosity in units of the Eddington luminosity;
Col. [5]: distance of the dissipation region from the black hole, in units of $10^{15}$ cm;
Col. [6]: size of the BLR, in units of $10^{15}$ cm;
Col. [7]: size of the torus, in units of $10^{15}$ cm;
Col. [8]: power injected in the jet in relativistic electrons, calculated in the comoving 
frame, in units of $10^{45}$ erg s$^{-1}$;
Col. [9]: magnetic field in G;  
Col. [10]: bulk Lorentz factor;
Col. [11]: viewing angle in degrees;
Col. [12]: relativistic Doppler factor.
% Col. [11] and Col. [12]: break and maximum Lorenz factor of the injected electron distribution;
% Col. [13] and Col. [14]: slopes of the injected electron distribution; 
% Col. [15]: logarithm of the total kinetic plus magnetic jet power, in erg s$^{-1}$.
% Col. [16]: logarithm of jet power in the form of radiation, in erg s$^{-1}$;
% The values of the powers and the energetics refer to {\it one} jet.
\label{para}}
\tablehead{\colhead{$z$} & \colhead{$M$} & \colhead{$L_{\rm d}$} & \colhead{$L_{\rm d}/L_{\rm Edd}$} & \colhead{$R_{\rm diss}$} & \colhead{$R_{\rm BLR}$} & \colhead{$R_{\rm torus}$} & \colhead{$P^\prime_{\rm e, jet, 45}$} & \colhead{$B$} & \colhead{$\Gamma$} & \colhead{$\theta_{\rm V}$} & \colhead{$\delta$}}
\colnumbers
%\begin{tabular}{l l l l l l l l l l l l l l l l l l}
%\hline
%\hline
%~$z$ &$M$ &$L_{\rm d}$ &$L_{\rm d}/L_{\rm Edd}$ &$R_{\rm diss}$ &$R_{\rm BLR}$ &$R_{\rm torus}$ &$P^\prime_{\rm e, jet, 45}$  &$B$ &$\Gamma$ &$\theta_{\rm V}$  \\ % &$\log P_{\rm r}$
%~[1] &[2] &[3] &[4] &[5] &[6] &[7] &[8] &[9] &[10] &[11] \\
% &[12] &[13] &[14]  &[15]  \\
%\hline
\startdata   
0.171   &3.4e8  &8.9e--4 &5e--4 &16.3  &81.6  &372  &9e--5 &0.32  &16.3 &2  &24.6 \\
%\hline
%\hline 
%\end{tabular}
%\vskip 0.4 true cm
\enddata
\end{deluxetable*}
% --------------------------------------

%%%%%%%%%%%%%%%%%%%%%%%%%%%%%%%%%%%%%%%%%%%%%%%%%%%%%%%%%%%%%%%%%%%%%%%%%%%
\section{Conclusions}

Blazars physics has received a tremendous boost from modern-era space missions like {\emph{Swift}}, {\emph{INTEGRAL}, and {\emph{Fermi}}. Many extragalactic sources detected in recent gamma-ray catalogues by \emph{Fermi} have been matched with known radio loud Blazars, and daily/weekly light curves has allowed to correlate in detail for the first time the variability of such sources, putting a milestone in the understanding of jet launching/collimation mechanism. A new intriguing class of objects is arising from radio/gamma correlation: radio-weak BL Lacs.
 
In this work we present the results of a radio/optical follow-up of a $\gamma$-ray flare for one of these objects. We can summarize the results as follows: 

1) The detected flat radio SED confirmed a BL Lac nature, with an orientation close to the line of sight, and did not show variability on a 4-months time window. A low radio power, as measured from the Rx factor, suggests that this object is at the boundaries between the radio-loud and radio-quiet populations. 

2) The disk luminosity in Eddington units (see tab. \ref{para}), is lower than the known typical values for RL and RQ AGN, as well as for Blazars (0.02 -- 0.2, \citealt{Ghisellini15}). In the light of the evolutionary track between flat spectrum radio quasars and BL Lacs proposed by \cite{Cavaliere}, this could suggest that radio-weak BL Lacs can experience flaring episodes in X/gamma-ray bands, despite the low accretion foreseen for objects at the end of the Blazar sequence.  

3) The intermediate properties between the FRI and BL Lac classes, indicated by the Ca$_{\rm II}$ H\&K break value, confirms the mixed state of this source. 

4) Finally, the overall SED fitting, from radio to $\gamma$-ray band, is typical of low-power, ``blue'' BL Lac.

From these results, we can conclude that the nature of this class of objects could probe the minimum conditions for the BL Lac class, both in terms of multi-wavelength variability and jet radio power. The lack of correlation in the explored time-window between the $\gamma$-ray band flux enhancement and radio activity in the GHz domain, could be symptomatic of a less efficient jet collimation, favoring a fast adiabatic expansion of the particle blobs responsible for the radio emission. This could smooth the additional contribution from newly injected particles. Another possibility is that the distance between the location of the $\gamma$-ray emitting region (i.e. the jet acceleration and collimation zone) and the mm-wave core (\citealt{Marscher}) is larger than in more powerful BL Lacs. Conversely, the blob speed could be lower, causing a further delay in the correlated radio flux enhancement.

The availability of future instruments like SKA in the radio band and CTA in $\gamma$-ray band, will make possible to discover and study in more detail populations of intermediate objects like this one, probing the minimum conditions for jet formation and collimation in the AGN population.

\acknowledgments
%%%%%%%%%%%%%%%%%%%%%%%%%%%%%%%% ACKNOWLEDGEMENTS %%%%%%%%%%%%%%%%%%%%%%%%%%%
We thank F. Tavecchio for useful discussions.
This publication has received funding from the European Union's Horizon 2020 research and innovation programme under grant agreement No. 730562 (RadioNet).
We acknowledge support from a grant PRIN-INAF SKA-CTA 2016.
GB acknowledges financial support under the INTEGRAL ASI-INAF agreement 2013-025.R01. 
VC and HAPH acknowledge support from CONACyT research grant 280789.
LHG acknowledges support from FONDECYT through grant 3170527.
Based on observations with the 100-m telescope of the MPIfR (Max-Planck-Institut f\"ur Radioastronomie) at Effelsberg.
Based upon observations carried out at the Observatorio Astron\'omico Nacional on the Sierra San Pedro M\'artir (OAN-SPM), Baja California, M\'exico.
Part of this work is based on archival data, software or online services 
provided by the ASI SCIENCE DATA CENTER (ASDC).
%

%%%%%%%%%%%%%%%%%%%%%%%%%%%%%%%%%%%%%%%%%%%%%%%%%%%%%%%%%%%%%%%%%%%%%%%%%%%

%%%%%%%%%%%%%%%%%%%%%%%%%%%%%%%%%%%%%%%%%%%%%%%%%%%%%%%%%%%%%%%%%%%%%%%%%%%


\begin{thebibliography}{}
 
\bibitem[\protect\citeauthoryear{Baars et al.}{1997}]{Baars} Baars, J. W. M., Genzel, R., Pauliny-Toth, I. I. K. et al. 1977, \aap , 61, 99

\bibitem[\protect\citeauthoryear{Becker et al.}{1995}]{Becker} Becker, R. H., White, R. L., \& Helfand, D. J. 1995, \apj, 450, 559

\bibitem[\protect\citeauthoryear{Bird et al.}{2016}]{Bird16} Bird, A. J., Bazzano, A., Malizia, A., Fiocchi et al. 2016, \apjs, 223, Issue 1, 15

\bibitem[\protect\citeauthoryear{Boccardi et al.}{2016}]{Boccardi16} Boccardi, B., Krichbaum, T. P., Bach, U. et al. 2016, \aap , 588, L9

\bibitem[\protect\citeauthoryear{Blandford \& Payne}{1982}]{Blandford} Blandford, R. D. \& Payne, D. G. 1982, \mnras, 199, 883

\bibitem[\protect\citeauthoryear{Bruzual \& Charlot}{2003}]{Bruzual} Bruzual, G. \& Charlot, M. J., 2003, MNRAS , 344, 1000

\bibitem[\protect\citeauthoryear{Cardelli et al.}{1989}]{Cardelli} Cardelli, J.A., Clayton, G.C., Mathis J. S. 1989, ApJ , 345, 245

\bibitem[\protect\citeauthoryear{Casadio et al.}{2015}]{CasadioCTA} Casadio, C.,  G\'omez, J. L., Jorstad, S. G. et al. 2015, ApJ, 813, 51

\bibitem[\protect\citeauthoryear{Cavaliere \& D'Elia}{2002}]{Cavaliere} Cavaliere, A. \& D'Elia, V. 2002, \apj, 571, 226

\bibitem[\protect\citeauthoryear{Chidiac et al.}{2016}]{Chidiac} Chidiac, C., Rani, B., Krichbaum, T. P. et al. 2016, A\&A, 590, A61

\bibitem[\protect\citeauthoryear{Chornock \& Margutti}{2017}]{Chornock} Chornock, R. \& Margutti, R. 2017, The Astronomer's Telegram, No. 10491

\bibitem[\protect\citeauthoryear{Cid Fernandes et al.}{2005}]{Cid05} Cid Fernandes, R., Mateus, A., Sodré, L. et al. 2005, MNRAS , 358, 363

\bibitem[\protect\citeauthoryear{Cid Fernandes et al.}{2007}]{Cid07} Cid Fernandes, R., Asari, N. V., Sodré, L. et al. 2007, MNRAS , 375, L16

\bibitem[\protect\citeauthoryear{Ciprini et al.}{2017}]{Ciprini17} Ciprini, S., Cheung, C. C., Kocevski, D. et al. 2017, The Astronomer's Telegram, No. 10482

\bibitem[\protect\citeauthoryear{Condon et al.}{1998}]{Condon} Condon, J. J., Cotton, W. D., Greisen, E. W. et al. 1998, \aj , 115, 5 

 \bibitem[\protect\citeauthoryear{Ghisellini \& Tavecchio}{2015}]{Ghisellini15} Ghisellini, G. \& Tavecchio, F. 2015, MNRAS, 448, 1060

\bibitem[\protect\citeauthoryear{Ghisellini \& Tavecchio}{2009}]{Ghisellini09} Ghisellini, G. \& Tavecchio, F. 2009, \mnras, 397, 985  % canonico %

\bibitem[\protect\citeauthoryear{Gregg et al.}{1996}]{Gregg96} Gregg, M.D., Becker, R.H., White, R.L. et al. 1996, \aj, 112, 407

\bibitem[\protect\citeauthoryear{Intema et al.}{2017}]{Intema17} Intema, H. T., Jagannathan, P., Mooley, K. P. et al. 2017, \aap, 598, A78

\bibitem[\protect\citeauthoryear{Karamanavis et al.}{2016}]{Karamanavis} Karamanavis, V., Fuhrmann, L., Krichbaum, T. P. et al. 2016, A\&A, 586, A60

\bibitem[\protect\citeauthoryear{Landt et al.}{2002}]{Landt} Landt, H., Padovani, P., Giommi, P. 2002, MNRAS, 336, 945

\bibitem[\protect\citeauthoryear{León-Tavares et al.}{2011}]{Leon} León-Tavares, J., Valtaoja, E., Chavushyan, V. H., et al. 2011, MNRAS, 411, 1127

\bibitem[\protect\citeauthoryear{Lisakov et al.}{2017}]{Lisakov} Lisakov, M. M., Kovalev, Y. Y., Savolainen, T. et al. 2017, MNRAS, 468, 4

 \bibitem[\protect\citeauthoryear{Marscher et al.}{2008}]{Marscher} Marscher, A.P., Jorstad, S.G., D'Arcangelo, F.D. et al. 2008, Nature, 452, 7190 

\bibitem[\protect\citeauthoryear{Massaro et al.}{2017}]{Massaro17} Massaro, F.,  Marchesini, E. J., D'Abrusco, R. et al. 2017, \apj , 834, 113 

\bibitem[\protect\citeauthoryear{Schinzel et al.}{2012}]{Schinzel} Schinzel, F. K., Lobanov, A. P., Taylor, G. B. et al. 2012, A\&A, 537, A70

\bibitem[\protect\citeauthoryear{Schlegel et al.}{1998}]{Schlegel} Schlegel, D.J., Finkbeiner, D. P., Davis, M., 1998, ApJ, 500, 525

\bibitem[\protect\citeauthoryear{Shen et al.}{2011}]{Shen} Shen, Y., Richards, G.T., Strauss, M.A. et al. 2011, \apjs, 194, 45

\bibitem[\protect\citeauthoryear{Silva et al.}{1998}]{Silva} Silva, L., Granato, G.L., Bressan, A. et al. 1998, \apj, 509, 103

\bibitem[\protect\citeauthoryear{Terashima \& Wilson}{2003}]{Terashima} Terashima, Y. \& Wilson, A.S. 2003, \apj, 583, 1

\bibitem[\protect\citeauthoryear{Tremaine et al.}{2002}]{Tremaine} Tremaine, S., Gebhardt, K., Bender, R. et al. 2002, ApJ, 574, 740




 

 
%%%%%%%%%%%TO BE ADDED%%%%%%%%%%%%
%Bruzual G., Charlot M. J., 2003, MNRAS , 344, 1000
%Cardelli J. A., Clayton G. C., Mathis J. S., 1989, ApJ , 345, 245
%Cid Fernandes R., Mateus A., Sodré L., Stasińska G., Gomes J. M., 2005, MNRAS , 358, 363
%Cid Fernandes R., Asari N. V., Sodré L., Stasińska G., Mateus A., Torres-Papaqui J. P., Schoenell W., 2007, MNRAS , 375, L16
%Landt, H., Padovani, P., & Giommi, P. 2002, MNRAS, 336, 945
%León-Tavares J., Valtaoja E., Chavushyan V. H., et al. 2011, MNRAS, 411, 1127
%Schlegel D. J., Finkbeiner, D. P., and Davis, M., 1998, ApJ, 500, 525 
%Tremaine S., et al. 2002, ApJ, 574, 740


\end{thebibliography}
\end{document}